\documentclass{svjour2}                    
\smartqed  
\usepackage{url}
\usepackage{graphicx}
\usepackage{color}

\usepackage{amssymb}
\usepackage{latexsym}
\usepackage{enumerate}
\usepackage{multirow}
\usepackage{twoopt}
\usepackage{array}
\usepackage{alltt}

\begin{document}
\title{Gaussian Networks Generated by Random Walks}

\author{Marco Alberto Javarone}
\institute{Marco Alberto Javarone \at
              Dept. of Mathematics and Computer Science - University of Cagliari, Cagliari, Italy\\
              DUMAS - Dept. of Human and Social Sciences - University of Sassari, Sassari, Italy\\
              \email{marcojavarone@gmail.com}}

\maketitle

\begin{abstract}
We propose a random walks based model to generate complex networks. Many authors studied and developed different methods and tools to analyze complex networks by random walk processes. Just to cite a few, random walks have been adopted to perform community detection, exploration tasks and to study temporal networks. Moreover, they have been used also to generate scale-free networks. 
In this work, we define a random walker that plays the role of ``edges-generator''. In particular, the random walker generates new connections and uses these ones to visit each node of a network.  
As result, the proposed model allows to achieve networks provided with a Gaussian degree distribution, and moreover, some features as the clustering coefficient and the assortativity show a critical behavior. 
Finally, we performed numerical simulations to study the behavior and the properties of the cited model.
\end{abstract}

\section{Introduction}\label{intro}
Complex networks constitute a powerful framework to study many real complex systems as the World Wide Web, social networks and brain networks~\cite{barabasi01}\cite{capocci01}\cite{sporns01}.
During last years, many models to generate complex networks and to study their properties have been developed.
For instance, models as the Barabasi-Albert model~\cite{barabasi02}, the Erd\"{o}s-Renyi graphs~\cite{erdos01} and the Watts-Strogatz model~\cite{watts01}, have attracted the attention of many scientists due to their usefulness to represent and to analyze systems belonging to different fields, e.g., biology, sociology, Internet and economy.
In particular, the Barabasi-Albert model is based on the preferential attachment mechanism and it allows to achieve scale-free networks, i.e., networks characterized by a degree distribution (i.e., the distribution of the number of connections) that follows a power-law~\cite{albert01}.
The Erd\"{o}s-Renyi graphs model generates networks with a binomial degree distribution, that converges to a Poissonian distribution under opportune conditions. These networks are also called classical random networks.
The Watts-Strogatz model implements an interpolation between an Erd\"{o}s-Renyi graph and a regular ring lattice, yielding networks with a homogeneous degree distribution.
Other models, to study the evolution of complex networks have been inspired from theoretical physics, for example the bosonic networks~\cite{bianconi01}\cite{bianconi02} and the fermionic networks~\cite{javarone01}. Both kinds of models aim to represent networks as quantum gases, showing that the networks evolution can be studied in terms of phase transitions --see also~\cite{hartonen01}.
Eventually, some authors studied the geometrical properties of complex networks. For instance, in the work of Krioukov et al.~\cite{krioukov01}, authors define a geometrical framework to generate networks characterized by a scale-free structure, by using a hyperbolic metrics.
In this work, we propose a method, based on random walk processes~\cite{lawler01}, to generate networks provided with a Gaussian degree distribution. Although Gaussian networks can be observed in some real systems, see for instance~\cite{chialvo01}\cite{strogatz01}, there are still not defined models to generate them. 
In the proposed model, we let a random walker to generate connections among nodes during its path. In particular, at each time step, the random walker visits a node and connects it with other nodes. In so doing, the random walker can jump from node to node by using these connections, choosing the destination node by a random selection among those available (i.e., the nodes connected to its position).
This dynamics can be viewed as a random walk process. In general, random walks have been widely used to analyze and to explore complex networks~\cite{zhou01}\cite{tadic01}\cite{starnini01}, both from a classical and from a quantum~\cite{biamonte01} computational perspective. Furthermore, in the work of Saram\"{a}ki and Kaski~\cite{saramaki01}, authors illustrate a mechanism, based on random walks, to generate scale-free networks.
We study the proposed model by numerical simulations, showing that it is possible to generate Gaussian networks and, moreover, that parameters as the clustering coefficient and the assortativity have a critical behavior.
The remainder of the paper is organized as follows: Section~\ref{sec:random-walker} briefly discusses random walk processes on complex networks and some applications. Section~\ref{sec:model_random_walker} introduces the proposed model based on random walk processes. Section~\ref{sec:results} shows results of numerical simulations. Finally, Section~\ref{sec:conclusions}) ends the paper.

\section{Random Walks on Complex Networks}\label{sec:random-walker}
In this section, we briefly introduce some basic concepts of random walks and we recall some applications to complex networks analysis.
In general, a random walk is a path composed by a sequence of random steps. It can be developed on different $n$-dimensional metric spaces. Probably, the most simple example of a $1$-dimensional random walk is the sequence of coin tosses. Formally, given a set of $1$-dimensional random variables $X$, whose values can be $+l$ or $-l$ (with probability $p$ and $1-p$, respectively) and a set $Y$, with $Y_0=0$ and $Y_n = \sum_{j=1}^{n} X_j$, the series $\{Y_n\}$ is called random walk. 
If the random walker is embedded in a network space, the random variables $X$ are the nodes of the network~\cite{noh01}. 
Let us consider a connected graph $G(V,E)$, with $V$ non-empty set of nodes (or vertices) and $E$ non-empty set of edges, represented by the adjacency matrix $A$. Elements $a_{ij}$, of the matrix $A$, represent the edges among nodes. In the case of a binary network, $a_{ij}= 1$ if nodes $i$ and $j$ are connected, otherwise $a_{ij} = 0$. 
Now, let us consider a random walker in the node $i$ at time $t$. At next step, i.e., at time $t+1$, it can jump to one of the $k_{i}$ neighbors of the node $i$ (with $k_i$ degree of the node $i$). In particular, the probability that it will jump to the node $j$ (at time $t+1$) is defined as:

\begin{equation} \label{eq:jump_probability}
p_{jump}(i \to j) = \frac{a_{ij}}{k_i}
\end{equation}

\noindent with $k_i$ computed as $k_i = \sum_z a_{iz}$. In general, if a random walker is in the node $i$ at $t=0$, the probability $P_{ij}$ to find it in the node $j$ at time $t$, is given by the master equation:

\begin{equation} \label{eq:stay_probability}
p_{ij}(t+1) = \sum_z \frac{a_{zj}}{k_z} p_{iz}(t)
\end{equation}
In order to analyze the dynamics of a random walker on complex networks, it is useful to estimate the number of steps required to cover a defined path, e.g., from a node $A$ to a node $B$. 
These measures are not trivial and different strategies can be developed to compute them (e.g.,~\cite{redner01}). In this context we are interested in the estimation of a parameter called ``cover time''~\cite{jonasson01} (we indicated as $Cov$). 
This parameter represents the number of steps used by the random walker to visit each node of a network.
In general, the value of $Cov$ depends on the topology of the random network. For instance, in a fully-connected network the cover time is computed as:
\begin{equation} \label{eq:cover_time_fully}
Cov = N ln(N)
\end{equation}
\noindent with $N$ number of nodes, whereas in Erd\"{o}s-Renyi graphs its value can be computed according to the following equation:
\begin{equation} \label{eq:cover_time}
Cov_{ER} = c \cdot ln(\frac{c}{c-1}) \cdot N ln(N)
\end{equation}
\noindent with $c >\ 1$, representing the connectedness of the network (i.e., there are not nodes with $k=0$).
\subsection*{Applications to network analysis}
Random walk processes can be used to analyze and to generate complex networks. For instance, they can be applied to community detection~\cite{newman01} as, if a network has a community structure, a random walker will spend long time inside each community because of the density of internal edges that they will be passed through.
In the work~\cite{tadic01}, the author shows a method to explore complex networks by using random walkers.
Moreover, in~\cite{hardiman01} authors develop a method to estimate the following properties of a network: the average clustering coefficient, the global clustering coefficient and the network size, i.e., the number of nodes. 
Finally, in the work~\cite{saramaki01}, it is shown a method to generate scale-free networks by using random walks. In particular, authors define the following algorithm:
\begin{enumerate}[1.]
\item The network is initialized with $m_0$ connected nodes.
\item The random walker is placed on a randomly chosen node.
\item At each step, the random walker jumps to a new node among the neighbors of the current node. After $l$ steps the node at which the random walker has arrived is marked. The random walker covers continuously a new path until $m \le m_0$ different nodes are marked.
\item A new node is added to the network and connected to the $m$ marked nodes; then the process is repeated from step ($2$) until the network has $N$ nodes.
\end{enumerate}
It is worth to recall that the above algorithm implements a preferential attachment mechanism.
\section{Gaussian Networks}\label{sec:model_random_walker}
\begin{figure}[!ht]
\centering
\includegraphics[width=5.0in]{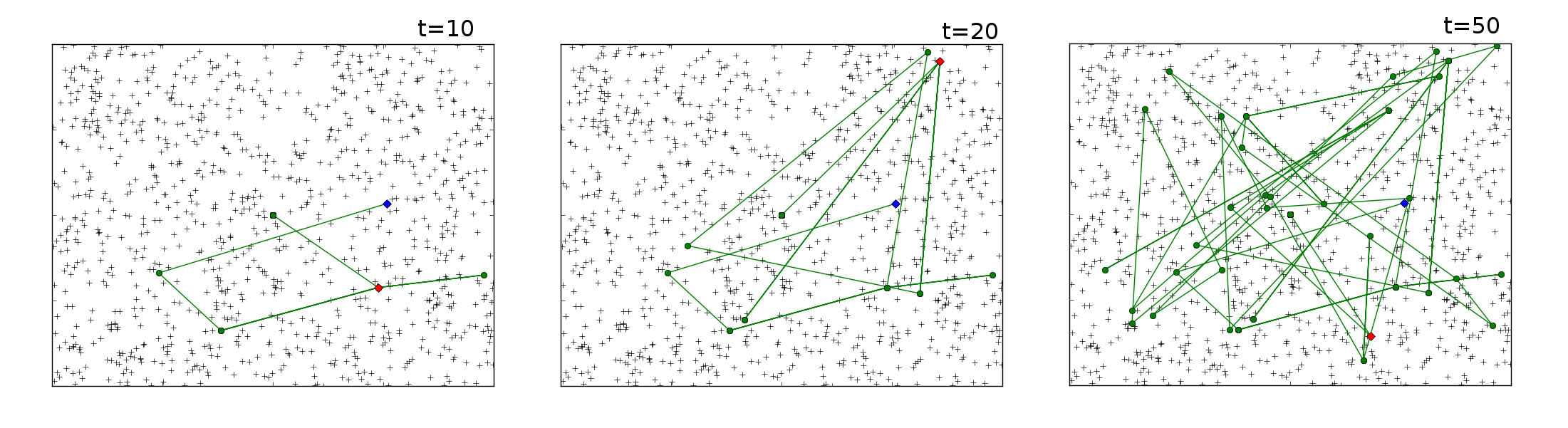}
\caption{Pathway of the random walker in a network with $N=1000$ and $m=3$. The blue diamond is the starting node, whereas the red diamond is the final node. The green points are visited nodes and the black crosses are unvisited nodes.}
\label{fig:random_path}
\end{figure}

\begin{figure}[!ht]
\centering
\includegraphics[width=3.5in]{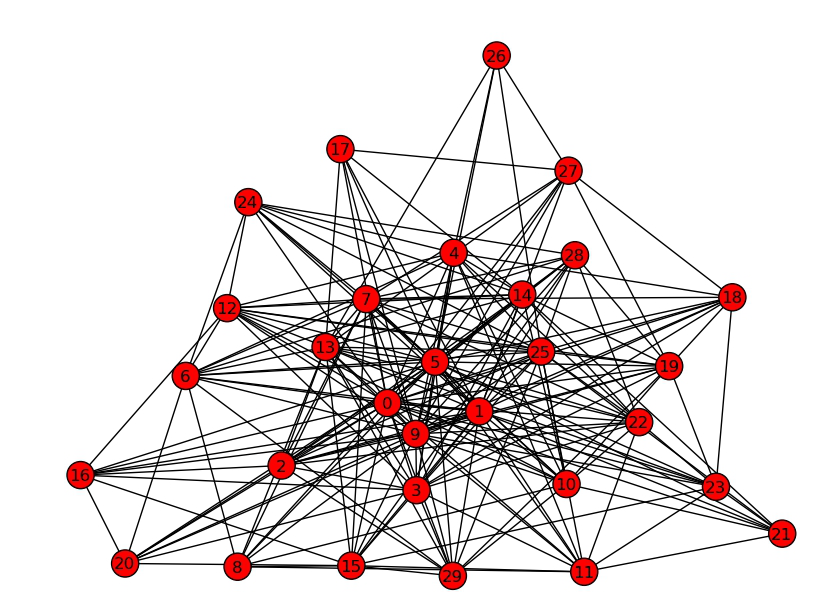}
\caption{A small network generated by the proposed method.}
\label{fig:small_network}
\end{figure}

Let us now to introduce a model, based on random walks, to generate Gaussian networks. We consider a set of $N$ nodes with an initial degree equal to zero, i.e., there are not edges in the network. A random walker, that wants to generate connections between nodes, is added to the system and it is placed on a randomly chosen node. 
At each time step, the random walker draws up to $m$ edges between the node on which it is located and other ones randomly chosen. Then, it jumps to one neighbor node, among those available from its position, selecting its destination ``$i$'' by Eq.~\ref{eq:jump_probability}:
\begin{itemize}
\item If the node $i$ is unvisited, it draws $m-k_i$ (with $k_i$ degree of the node $i$) new edges between the node $i$ and other randomly chosen nodes.
\item Else it draws only one new edge between the node $i$ and another randomly chosen node.
\end{itemize}
Note that, if the node $i$ is unvisited and $k_i \ge m$ the random walker cannot generate any new edge, therefore it can just select a new node to visit. 
Instead, if the node $i$ has been already visited, the random walker draws a new edge to avoid infinite loop in the algorithm, i.e., to avoid to be trapped in a small set of nodes with a degree $ k \ge m$.
Obviously, in both cases, connections are generated only between two nodes, e.g., $i$ and $j$, that are still not connected (i.e., $a_{ij}=0$).
The algorithm ends when all nodes have been visited by the random walker.
To summarize, the proposed method is composed by the following steps:
\begin{enumerate}[1.]
\item Define $N$ nodes and $m$ maximum number of new edges generated at each time step.
\item Place a random walker on the $x$th node (randomly chosen).
\item The random walker connects the $x$th node with $m$ (randomly selected) nodes.
\item The random walker jumps to a neighbor node (selected according to Eq.~\ref{eq:jump_probability}).
\item The random walker verifies whether its destination node ``$i$'' is unvisited:
\begin{enumerate}[i.]
\item IF ``$i$'' is unvisited, the random walker connects it with $m-k_i$ nodes;
\item ELSE the random walker connects it with only one node.
\end{enumerate}
\item Repeat from $(4)$ until all nodes have been visited by the random walker.
\end{enumerate}
Figure~\ref{fig:random_path} illustrates three screenshots of the pathway covered by the random walker in a small network ($N=1000$, $m=3$).
Although, the random walker moves among nodes following specific directions, i.e., from the node $x$ to the node $y$, it is possible to generate both directed and undirected networks. 
Figure~\ref{fig:small_network} shows a small networks ($N=30, m=2$) generated by the proposed model.
There are many differences between our model and that described in~\cite{saramaki01}. Just to cite a few, we consider an initial set of $N$ unconnected nodes and a random walker that generates edges at each time step. Moreover, the proposed model does not yield scale-free networks, but networks provided with a Gaussian degree distribution. 
\section{Simulations}\label{sec:results}
We performed many numerical simulations to study the proposed model. In particular, we analyzed the main properties of simulated networks, as the degree distribution, the path length, the clustering coefficient~\cite{watts01} and the assortativity~\cite{newman02}. 
Simulations have been performed with a number of nodes $N$ in the range $[1000, 50000]$ and values of $m$ in the range $[1,50]$.
\subsection*{Degree distribution}
\begin{figure}[!ht]
\centering
\includegraphics[width=5.0in]{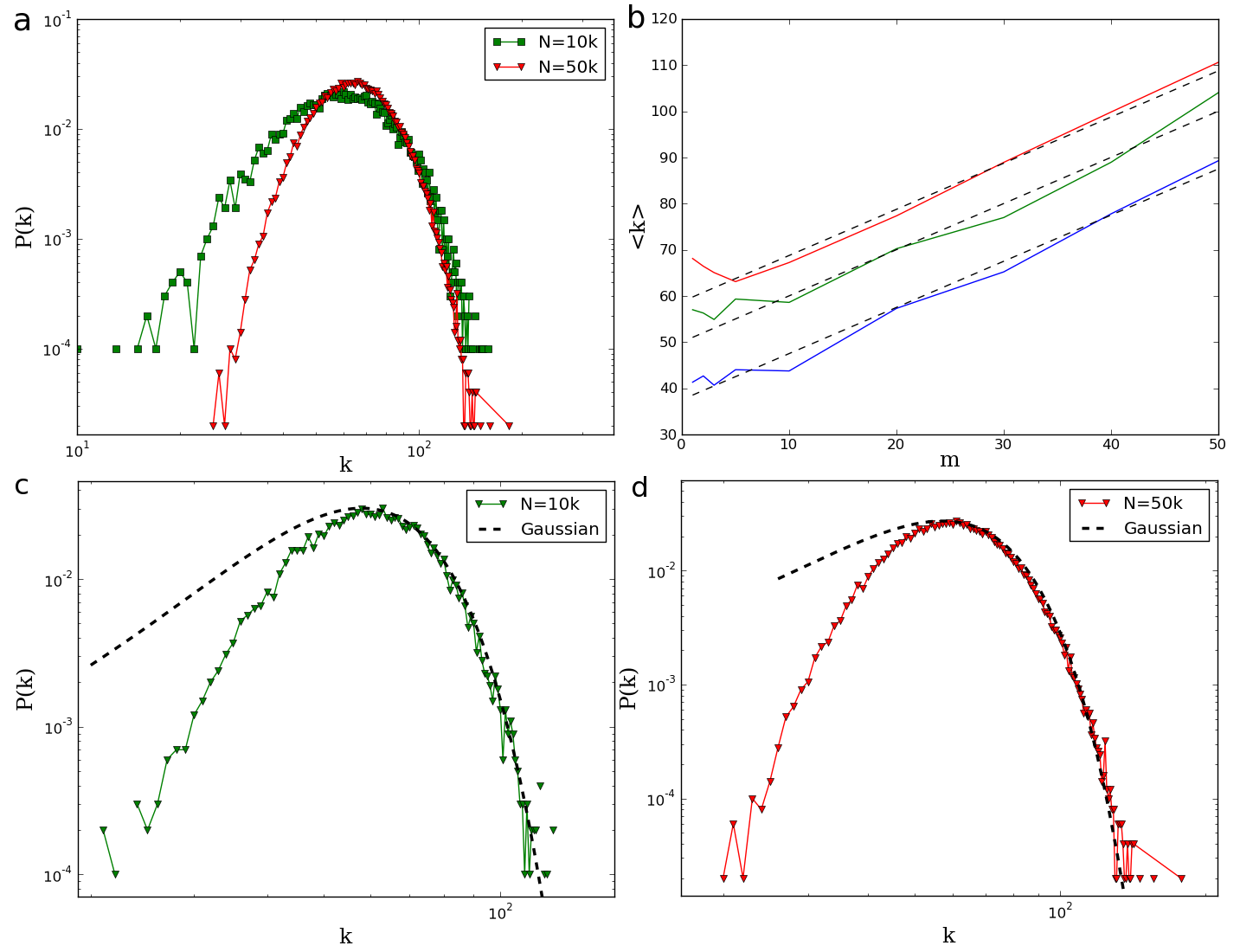}
\caption{\textbf{a} Networks of different size, as indicated in the legend, generated by the proposed modle with $m=5$. \textbf{b} Average degree distribution achieved varying the value of $m$. Each curve refers to networks of different size. The black dotted lines fit the achieved results. Results are averaged over $20$ different realizations. \textbf{c} Degree distribution (green line) of a network with $N=10000$ and $m=10$, fitted by a Gaussian function (black dotted line). \textbf{d} Degree distribution (red line) of a network with $N=50000$ and $m=10$, fitted by a Gaussian function (black dotted line).}
\label{fig:degree_distribution}
\end{figure}

The first analysis is related to the studying of the degree distribution $P(k)$ of simulated networks --see Panel~\textbf{a} of Figure~\ref{fig:degree_distribution}. 
In general, we found that for small values of $m$ networks are characterized by the presence of a few hubs, since as $m$ increases the right tail of the $P(k)$ reduces (considering networks with the same size). 
In the proposed model, the analytical definition of the average degree $\langle k \rangle$ is not trivial.
Notwithstanding, by considering the dynamics of the random walker, we derive the following equation:
\begin{equation} \label{eq:average_degree}
\langle k \rangle \sim m + \frac{Cov_{ER}}{N}
\end{equation}
\noindent with $Cov_{ER}$ defined in Eq.~\ref{eq:cover_time}. 
In particular, the degree of each node is equal to the sum of $m$ with the amount of further visits, as at least a new link is generated each time. 
To estimate the number of further visits to each node, we consider the structure of a growing classical random network (by using $Cov_{ER}$).
Values of $\langle k \rangle$, achieved in simulated networks, allow to compute the parameter $c\sim 1,005$. In Panel~\textbf{b} of Figure~\ref{fig:degree_distribution} it is possible to compare the results achieved in simulated networks with the theoretical values computed by Eq.~\ref{eq:average_degree}.
Finally, as shown in Panels~\textbf{c-d} of Figure~\ref{fig:degree_distribution}, we found that Gaussian functions allow to fit the achieved degree distributions. Therefore, we call these networks ``Gaussian Networks''.
\subsection*{Path Length and Clustering Coefficient}
\begin{figure}[!ht]
\centering
\includegraphics[width=4.5in]{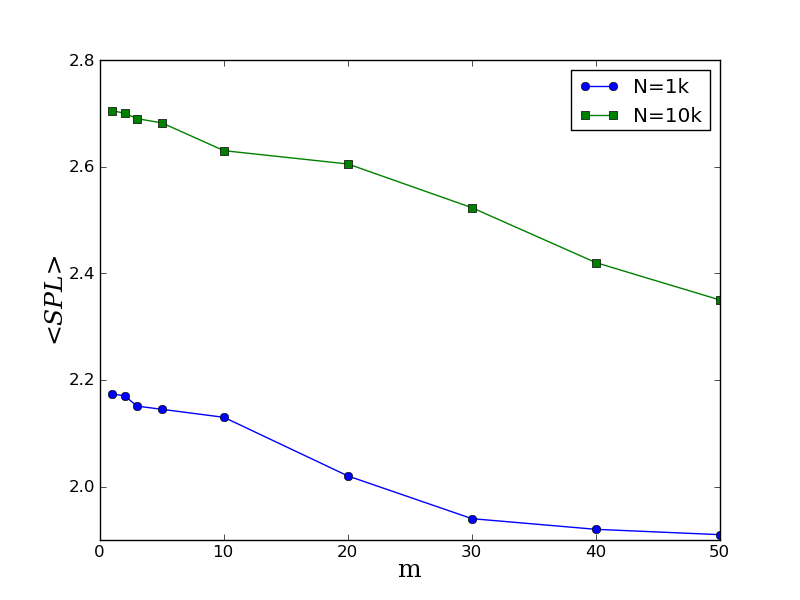}
\caption{Shortest Path Lengths (\textit{SPL}) computed in networks with $N=1000$ and $N=10000$, varying the value of $m$. Results are averaged over $20$ different realizations.}
\label{fig:path_length}
\end{figure}

\begin{figure}[!ht]
\centering
\includegraphics[width=5.0in]{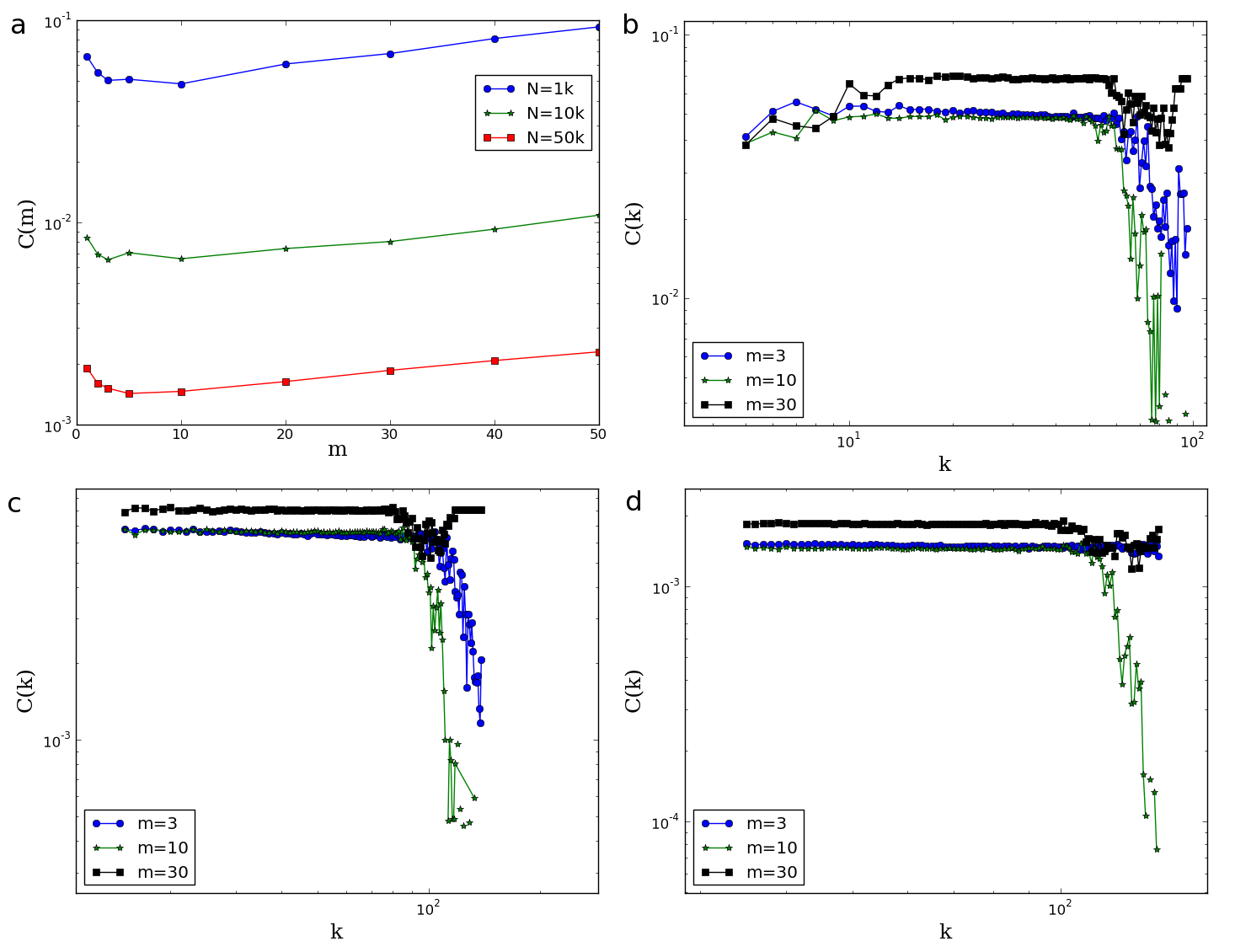}
\caption{Clustering coefficients computed in networks generated by random walks. \textbf{a} Average clustering coefficient in function of $m$. \textbf{b} Clustering coefficient in function of the node degree in networks with $N = 1000$. \textbf{c} Clustering coefficient in function of the node degree in networks with $N = 10000$. \textbf{d} Clustering coefficient in function of the node degree in networks with $N = 50000$. Results are averaged over $20$ different realizations.}
\label{fig:clustering}
\end{figure}

The analysis of the path length allows to evaluate whether simulated networks show a small-world behavior. 
The latter implies that the distance between two randomly chosen nodes, called also average shortest path lenght (or $\langle SPL \rangle$), grows proportionally to the logarithm of the network size, i.e., $\langle SPL \rangle \sim \ln(N)$ --see ~\cite{watts01}.
In Figure~\ref{fig:path_length}, we show results achieved during numerical simulations. 
As expected, the $\langle SPL \rangle$ depends on the network size $N$ and, moreover, it decreases as the value of $m$ increases. 
In particular, as the number of edges (i.e., $m$), generated at each time step, increases the average shortest distance decreases, since more paths are available.
It is interesting to note that values of $\langle SPL \rangle$ do not scale logarithmically with the network size (i.e.,  with $\ln(N)$), but they scale with $\ln \ln (N)$.
Therefore, networks generated by the proposed model can be defined as ``ultrasmall'' --see ~\cite{cohen01}.
\newline
Now, we consider the clustering coefficient $C$ computed in networks simulated varying both $N$ and $m$. In general, the clustering coefficient allows to know if nodes of a network tend to cluster together, therefore this measure supplies a further information on the network structure.
At the beginning, we consider the average value of $C$ as a function of $m$. Related results are shown in the panel~\textbf{a} of Figure~\ref{fig:clustering}. 
It is interesting to note a strange behavior of the considered parameter $C(m)$ for low values of $m$. In particular, starting from $m=1$ and increasing up to $m=5$, the value of $C(m)$ decreases. Then, as $m$ increases from $5$ to the $50$, $C(m)$ linearly increases.
We also analyze the variation of the clustering coefficient with the node degree (i.e., $C(k)$) --see panels~\textbf{b-c-d} of Figure~\ref{fig:clustering}.
Results of this analysis show that the clustering coefficient is higher for nodes with a small degrees than for those with a big degree. Furthermore, this parameter has a steady-state behavior until a (relatively) high $k$, where we observe a breaking point of such behavior, followed by a fast exponential decrease of $C$ as $m\sim 10$.  
\subsection*{Assortativity}
\begin{figure}[!ht]
\centering
\includegraphics[width=4.5in]{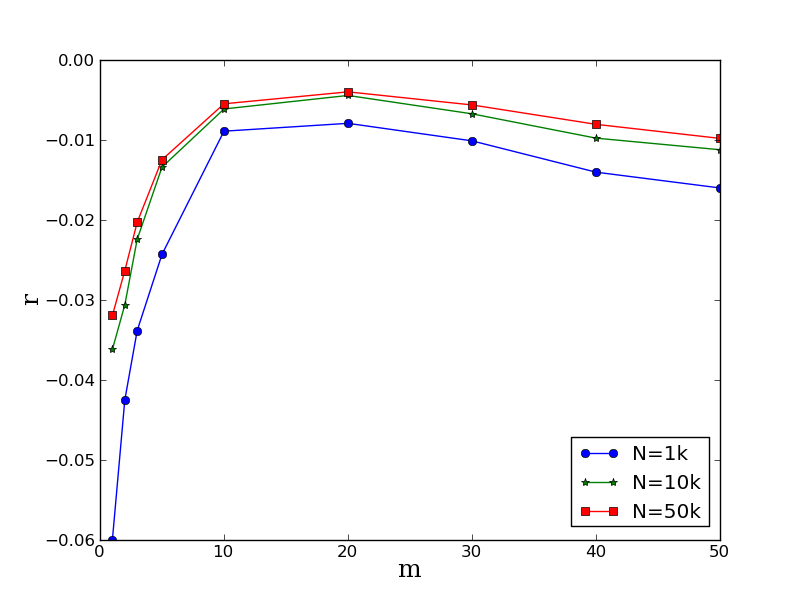}
\caption{Assortativity in networks of different size. Results are averaged over $20$ different realizations.}
\label{fig:assortativity}
\end{figure}

Eventually, we analyze the assortativity of simulated networks. The assortativity is a property that allows to evaluate if nodes prefer to attach to other nodes that are (not) similar. 
This property affects the whole structure of a network. For instance, social networks can be divided into communities of users speaking the same language or having same hobbies. 
A network is assortative when its assortativity $r$ is positive and, on the contrary, it is disassortative when $r$ is negative.
The similarity between nodes can refers to their degree. In this case, Johnson et al.~\cite{torres01} found a relation between the assortativity and the Shannon entropy of networks. In particular, they demonstrated that scale-free networks have a high probability to be disassortative.
Results of this analysis illustrate that the assortativity strongly depends on $m$ --see Figure~\ref{fig:assortativity}.
Although all simulated networks are disassortative (i.e., $r <\ 0$), it is worth to observe that $r$ increases almost up to $0$ as $m$ increases from $m=1$ to $m\sim20$, then this value slowly decreases as $m$ increases. Moreover, this behavior appears in networks of different size, i.e., it does not depend on $N$.

\section{Discussion and Conclusions}\label{sec:conclusions}
In this work, we propose a model to generate complex networks by using a random walker. As reported in previous works, random walks can be used to analyze complex networks, e.g., performing community detection or exploration tasks and also to generate scale-free networks~\cite{saramaki01}. 
We implement a random walker that plays the role of ``edges-generator''. As shown by numerical simulations, the cited model allows to achieve networks characterized by a Gaussian degree distribution. Networks provided with this distribution can be observed in some real systems, as described in~\cite{chialvo01}\cite{strogatz01}. 
Notwithstanding, there are not models for generating networks provided with this kind of degree distribution.
Simulated networks allow to measure important parameters as the average shortest path length, the clustering coefficient and the assortativity.
The analysis of the path lenght shows that these networks are ``ultrasmall'' (see~\cite{cohen01}), as their average shortest path length scales as $\ln \ln N$.
The clustering coefficient and the assortativity have a critical behavior in networks generated with small values of $m$. In particular, the average value of the former has an initial decrease as $m$ increases from $m=1$ to $m= 5$. 
After this value, the average $C$ linearly increases with $m$.
Moreover, we observe a fast transition from (relatively) high values of $C$ (as a function of the node degree $k$) to (relatively) low values, by using $m\sim10$. In particular, the function $C(k)$ has a steady-state behavior until $k$ reaches a critical (relatively) high value.
On the other hand, the assortavity $r$ of simulated networks is always negative, i.e., they are disassortative. As before, it is worth to observe a difference, in the behavior of the proposed model, between results achieved with small $m$ and those achieved with big $m$. 
In particular, as $m$ increases from $m=1$ to $m=\sim20$, the value of $r$ increases. Then, $r$ slowly decreases as $m$ further increases.
We deem that the critical behavior of the proposed model strictly depends on the rule followed by the random walker to generate the edges. For small values of $m$ the random walker spends more time to generate all the necessary edges, whereas for high values of $m$ in a few time steps many nodes are immediately connected.
Furthermore, the rule that allows the random walker to generate one more edge, in the event it visits an already visited node, is responsible for the formation of hubs (i.e., nodes provided with a high degree).
Eventually, the proposed model represents a new approach to generate random networks, provided with well defined features as the Gaussian degree distribution, the ``ultrasmall'' behavior, and the negative assortativity. 
In particular, it is worth to note that the ``ultrasmall'' behavior has been observed in scale-free networks, that are much different from those illustrated in this work.
Therefore, we deem the proposed model useful to study networks characterized by the described parameters, since these ones can be found in real networks and they can also be observed in different networks provided with more popular structures (e.g., scale-free).
\section*{Acknowledgments}
The author would like to thank Fondazione Banco di Sardegna for supporting his work.

\end{document}